\documentstyle[aps,pre,epsf]{revtex}

\begin{document}
\draft 
\title{Medium-range interactions and crossover to classical critical behavior}
\author{Erik Luijten\cite{email} and Henk W. J. Bl\"ote}
\address{Department of Physics, Delft University of Technology, P.O. Box
  5046, 2600 GA Delft, The Netherlands}
\author{Kurt Binder}
\address{Institut f\"ur Physik, Johannes Gutenberg Universit\"at, Staudinger
  Weg 7, D-55128 Mainz, Germany}

\date{\today}
\maketitle
\begin{abstract}
  We study the crossover from Ising-like to classical critical behavior as a
  function of the range $R$ of interactions. The power-law dependence on $R$ of
  several critical amplitudes is calculated from renormalization theory.  The
  results confirm the predictions of Mon and Binder, which were obtained from
  phenomenological scaling arguments. In addition, we calculate the range
  dependence of several corrections to scaling.  We have tested the results in
  Monte Carlo simulations of two-dimensional systems with an extended range of
  interaction.  An efficient Monte Carlo algorithm enabled us to carry out
  simulations for sufficiently large values of $R$, so that the theoretical
  predictions could actually be observed.
\end{abstract}


\section{Introduction}
As is well known, the critical behavior of a physical system strongly depends
on the range of the interactions. The longer the range, the stronger critical
fluctuations will be suppressed. In the limit of infinite range we recover
classical or mean-field-like critical behavior.  For interactions with a finite
range, however, fluctuations remain very important and essentially modify the
critical behavior. As follows from the Ginzburg criterion~\cite{ginzburg},
sufficiently close to the critical temperature $T_{\rm c}$ nonclassical
critical exponents apply for any finite interaction range $R$.  This crucial
difference between finite and infinite $R$ implies a crossover from one type of
critical behavior to another as a function of $R$.  Such crossover phenomena
are of great interest for a wealth of critical phenomena. They occur, e.g., in
polymer mixtures (see Ref.~\cite{deutsch} and references therein), as a
function of the chain length, and gas--liquid transitions, as a function of the
difference between the temperature and the critical temperature. The
explanation of these phenomena in terms of competing fixed points of a
renormalization transformation is one of the important features of the
renormalization theory (see, e.g., Ref.~\cite{rg-review}).

In Ref.~\cite{monbin}, Mon and Binder have already studied crossover as a
function of $R$ within the context of finite-size scaling, motivated by the
crossover in polymer mixtures. They predicted that the critical amplitudes of
scaling functions display a singular dependence on $R$. The various power-law
dependencies were obtained from phenomenological crossover scaling arguments.
In this paper, we will derive this dependence on $R$ from a renormalization
description of the crossover from classical to nonclassical critical behavior.
The first part of the renormalization trajectory is governed by the Gaussian
fixed point, which is unstable for $d<4$. The corresponding scaling relations
have been derived by Rikvold et al.~\cite{rikvold}. Sufficiently close to
criticality, the final part of the renormalization trajectory is governed by
the Ising fixed point.  The resulting relations are in complete agreement with
the predictions from Ref.~\cite{monbin}. In addition, we obtain the $R$
dependence of the leading corrections to scaling and derive from
renormalization arguments a logarithmic factor in the shift of the critical
temperature. This factor was already conjectured in Ref.~\cite{monbin}.

It is interesting to note that the physical mechanism leading to the singular
range dependence of scaling functions is closely related to that leading to the
violation of hyperscaling for $d>4$. The latter effect is caused by a singular
dependence of thermodynamic quantities on the coefficient $u$ of the $\phi^4$
term for $u \to 0$ in a Landau-Ginzburg-Wilson Hamiltonian. In other words, $u$
is a so-called dangerous irrelevant variable (see, e.g., Ref.~\cite{univers}
for a more detailed discussion). Here, as we will see, the fact that this
coefficient becomes small for large values of $R$ plays again an essential
role, although $u$ is relevant for $d<4$.

Furthermore, we present new Monte Carlo results for two-dimensional Ising
models with an extended range of interaction. A serious problem associated with
such simulations is that the simulation time tends to increase rapidly with the
number of interacting spins. However, a large interaction range is crucial to
observe the predicted $R$ dependencies, as will follow from the renormalization
description. The maximum range that could be accessed in Ref.~\cite{monbin} was
too small to verify the theoretical predictions.  We overcome this limitation
by means of an efficient cluster Monte Carlo algorithm~\cite{ijmpc}, in which
the simulation time per spin is practically independent of the range of the
interactions.

The outline of this paper is as follows. In section~\ref{sec:theory}, we derive
the $R$ dependence of critical amplitudes from renormalization theory. These
results are verified by Monte Carlo simulations, presented in
section~\ref{sec:mc}. We end with our conclusion in section~\ref{sec:concl}.
Two technical issues, namely the Fourier transform of a spherical distribution
of interactions in a general number of dimensions and the application of the
cluster algorithm to medium-range interactions, are addressed in Appendices
\ref{sec:fourier} and~\ref{sec:algorithm}, respectively.

\section{Renormalization derivation of the dependence of critical amplitudes
 on the interaction range}
\label{sec:theory}
A model with long-range interactions which has attracted much attention is that
in which the spin--spin interactions decay algebraically as a function of the
distance $r$ between the spins, $K(r)=Ar^{-d-\sigma}$ ($\sigma > 0$), where $d$
is the dimensionality and $A>0$ the interaction strength (see, e.g.,
Refs.~\cite{lr-rg,sak73,univers}). For $d<4$, this model displays an
interesting continuous variation of critical behavior as function of $\sigma$:
for $0 < \sigma \leq d/2$ the critical behavior is classical (mean-field-like),
for $d/2 < \sigma < 2-\eta_{\rm sr}$ the critical exponents vary continuously,
and for $2-\eta_{\rm sr} < \sigma$, the interactions decay fast enough to yield
short-range (Ising-like) critical behavior.  Here, $\eta_{\rm sr}$ denotes the
exponent $\eta$ in the $d$-dimensional model with short-range interactions. The
algebraical decay of the interactions is responsible for the existence of an
intermediate regime between Ising-like and classical critical behavior.  In
this paper we focus on a different way to interpolate between the long-range
(mean-field) limit and short-range models.  Instead, we choose ferromagnetic
interactions which are constant within a range $R$ and zero beyond this range.
Thus, we have the following Hamiltonian
\begin{equation}
{\cal H}/k_{\rm B}T = - \sum_i \sum_{j > i} K_d({\bf r}_i-{\bf r}_j) s_i s_j
                      - h_0 \sum_i s_i \;,
\label{eq:latham}
\end{equation}
where the spin--spin interaction $K_d({\bf r}) \equiv cR^{-d}$ for $|{\bf r}|
\leq R$ and the sums run over all spins in the system. This Hamiltonian
displays physical behavior that is different from the power-law case. In
particular, the intermediate regime with variable exponents is absent, and
mean-field critical behavior is restricted to the infinite-range limit. We
analyze the influence of the range $R$ within the context of renormalization
theory, starting from a generalized Landau-Ginzburg-Wilson (LGW) Hamiltonian,
where the $[\nabla \phi({\bf r})]^2$ term normally representing the
(short-range) interactions is replaced by an interaction term with spin--spin
coupling~(\ref{eq:cstint}),
\begin{equation}
  {\cal H}(\phi)/k_{\rm B}T = \int_{V} d{\bf r}  \left\{ - \frac{1}{2}
  \int_{|{\bf r}-{\bf r}'| \leq R} d{\bf r}' \left[ \frac{c}{R^d} \phi({\bf r})
  \phi({\bf r}') \right] - h_0\phi({\bf r}) + \frac{1}{2} v \phi^2({\bf r})
  + u_0 \phi^4({\bf r}) \right\} \;.
\label{eq:hamil}
\end{equation}
As a consequence of the normalization factor $R^{-d}$, the critical value of
the temperature parameter $c$ depends only weakly on $R$. The first integral
runs over the volume $V$ which contains $N$ particles. We adopt periodic
boundary conditions.  The Fourier transform of the interaction is calculated in
Appendix~\ref{sec:fourier}. It leads to the following momentum-space
representation of the Hamiltonian
\begin{equation}
  {\cal H}(\phi_{\bf k})/k_{\rm B}T = \frac{1}{2} \sum_{{\bf k}} \left[
  - c \left(\frac{2\pi}{kR}\right)^{d/2} J_{d/2}(kR) + v \right]
  \phi_{\bf k} \phi_{-{\bf k}}  +  \frac{u_0}{4N} \sum_{{\bf k}_1} \sum_{{\bf
  k}_2} \sum_{{\bf k}_3} \phi_{{\bf k}_1} \phi_{{\bf k}_2} \phi_{{\bf k}_3}
  \phi_{-{\bf k}_1-{\bf k}_2-{\bf k}_3} - h_0\sqrt{\frac{N}{2}} \phi_{{\bf
  k}={\bf 0}} \;.
\label{eq:fourier}
\end{equation}
$J_\nu$ is a Bessel function of the first kind of order $\nu$. The wavevectors
are discrete because of the periodic boundary conditions. Furthermore, we
restrict the wavevectors to lie within the first Brillouin zone, which is
reminiscent of the underlying lattice structure.  The interaction term can be
expanded in a Taylor series containing only even terms in $kR$.  This means
that we will be mainly concerned with the term of order $(kR)^2$, because
higher-order terms will turn out to be irrelevant. The constant term in the
Taylor series is absorbed in $\bar{v}$ and the coefficient of the quadratic
term as a factor in $\bar{c}$.  This yields a new Hamiltonian
\begin{equation}
  {\cal H}_{\rm t}(\phi_{\bf k})/k_{\rm B}T = \frac{1}{2} \sum_{{\bf k}} \left[
  \bar{c} R^2 k^2 + \bar{v} \right]
  \phi_{\bf k} \phi_{-{\bf k}}  +  \frac{u_0}{4N} \sum_{{\bf k}_1} \sum_{{\bf
  k}_2} \sum_{{\bf k}_3} \phi_{{\bf k}_1} \phi_{{\bf k}_2} \phi_{{\bf k}_3}
  \phi_{-{\bf k}_1-{\bf k}_2-{\bf k}_3} - h_0\sqrt{\frac{N}{2}} \phi_{{\bf
  k}={\bf 0}} \;.
\label{eq:trunc}
\end{equation}
Since we are free to choose the scale on which the fluctuations of the order
parameter are measured, we may rescale $\phi \to \psi \equiv \sqrt{\bar{c}}R
\phi$.  This is generally the most convenient choice because the dominant
$k$-dependent term becomes independent of $R$.\footnote{Naturally, this
  rescaling is not compulsory and the same results will be obtained without it,
  provided one keeps track of the dependence of the nontrivial fixed point on
  $\bar{c}R^2$, arising from the integration over propagators in the inner part
  of the Brillouin zone.} This leads to
\begin{equation}
  \tilde{{\cal H}}(\psi_{\bf k})/k_{\rm B}T = \frac{1}{2} \sum_{{\bf k}} \left[
  k^2 + \frac{\bar{v}}{\bar{c} R^2} \right]
  \psi_{\bf k} \psi_{-{\bf k}}  +  \frac{u_0}{4 \bar{c}^2 R^4 N}
  \sum_{{\bf k}_1} \sum_{{\bf k}_2} \sum_{{\bf k}_3}
  \psi_{{\bf k}_1} \psi_{{\bf k}_2} \psi_{{\bf k}_3}
  \psi_{-{\bf k}_1-{\bf k}_2-{\bf k}_3} -
  \frac{h_0}{\sqrt{\bar{c}}R} \sqrt{\frac{N}{2}} \psi_{{\bf k}={\bf 0}} \;.
\label{eq:scal-hamil}
\end{equation}
The parameter $\bar{c}$ is merely a constant, {\em independent of the range},
and in order not to be hampered by it in the future analysis, we absorb the
various powers of it in $r_0 \equiv \bar{v}/\bar{c}$, $u \equiv u_0/\bar{c}^2$,
and $h \equiv h_0/\sqrt{\bar{c}}$. Now, $r_0$ assumes the role of the
temperature parameter.

\begin{figure}
\begin{center}
\leavevmode
\epsfbox{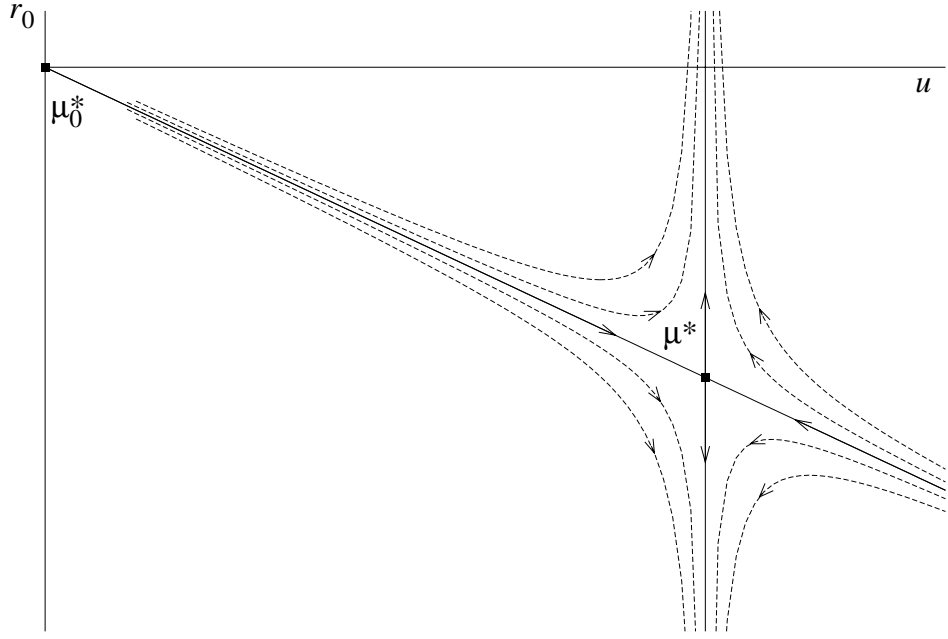} 
\end{center}
\caption{Qualitative picture of the renormalization trajectory describing the
  crossover from the Gaussian fixed point $\mu_0^* = (0,0)$ to the Ising fixed
  point $\mu^*=(r_0^*,u^*)$.}
\label{fig:traject}
\end{figure}

If the range $R$ is large, the coefficient of the $\psi^4$ term is relatively
small and hence the critical behavior of the system is determined by the
Gaussian fixed point. Under a renormalization transformation with a rescaling
parameter $l$ the Hamiltonian thus transforms as
\begin{eqnarray}
  \tilde{{\cal H}}'(\psi'_{{\bf k}'})/k_{\rm B}T' &=& \frac{1}{2} \sum_{{{\bf
  k}'}} \left[ k'^2 + \frac{r_0}{R^2}l^2 \right]
  \psi'_{{\bf k}'} \psi'_{-{\bf k}'}  +  \frac{u}{4 R^4 N'}l^{4-d}
  \sum_{{\bf k}_1'} \sum_{{\bf k}_2'} \sum_{{\bf k}_3'}
  \psi'_{{\bf k}_1'} \psi'_{{\bf k}_2'} \psi'_{{\bf k}_3'}
  \psi'_{-{\bf k}_1'-{\bf k}_2'-{\bf k}_3'} \nonumber \\ && -
  \frac{h}{R} \sqrt{\frac{N'}{2}}l^{1+d/2} \psi'_{{\bf k}'={\bf 0}} \;.
\label{eq:rg-step1}
\end{eqnarray}
Here $\psi'_{{\bf k}'}=l^{-1}\psi_{\bf k}$, ${\bf k}'={\bf k}l$, the sums run
again over the full Brillouin zone, and $N'=Nl^{-d}$.  For $d<4$, the $\psi^4$
term grows and the system moves away from the Gaussian fixed point $\mu_0^{*}$
(see Fig.~\ref{fig:traject}).  At present, we are interested in the flow from
the neighborhood of the Gaussian fixed point to that of the Ising fixed point.
Thus we remain close to the critical line connecting the two fixed points and the
temperature field parametrized by $r_0$ remains small.  The crossover to
Ising-like critical behavior occurs when the coefficient of the $\psi^4$ term
is of the same order as that of the $k^2\psi^2$ term, which is unity,
i.e., when $l = l_0 \equiv R^{4/(4-d)}$. We shall refer to $l_0$ as the {\em
crossover scale}.

By comparing the coefficient of the $\psi^4$ term to that of the $r_0\psi^2$
term, it is possible to derive a criterion that states for which temperatures
the critical behavior will be Ising-like and for which temperatures it will be
classical.  This is the well-known Ginzburg criterion~\cite{ginzburg}, which
can also be derived from Eq.~(\ref{eq:rg-step1}) (see, e.g.,
Ref.~\cite[p.~107]{fisher-sa}).  One expects the Gaussian fixed point to
dominate the renormalization flow if, irrespective of $l$, the $\psi^4$
coefficient is small compared to the temperature coefficient. Thus, one
requires the scaled combination $uR^{-4}l^{4-d} / (r_0R^{-2}l^2)^{(4-d)/2}$ to
be small, or $r_0^{(4-d)/2}R^d u^{-1} \gg 1$ (cf.\ also
Ref.~\cite[Eq.~(3)]{monbin}).

Since we are now in the neighborhood of the Ising fixed point, we continue
renormalizing our Hamiltonian with {\em nonclassical\/} renormalization
exponents $y_{\rm t}$, $y_{\rm h}$, and $y_{\rm i}$. To leading order, it will
transform as follows, where $b$ denotes the rescaling factor of our new
transformation,
\begin{eqnarray}
  \tilde{{\cal H}}''(\psi''_{{\bf k}''})/k_{\rm B}T''
  &=& \frac{1}{2} \sum_{{\bf
  k}''} \left[ k''^2 + R^{2d/(4-d)} (b^{y_{\rm t}}\tilde{r}_0 + r_0^*) \right]
  \psi''_{{\bf k}''} \psi''_{{\bf -k}''}  +
  \frac{b^{y_{\rm i}}\tilde{u} + u^* }{4 N''}
  \sum_{{\bf k}_1''} \sum_{{\bf k}_2''} \sum_{{\bf k}_3''}
  \psi''_{{\bf k}_1''} \psi''_{{\bf k}_2''} \psi''_{{\bf k}_3''}
  \psi''_{-{\bf k}_1''-{\bf k}_2''-{\bf k}_3''} \nonumber \\
  && - h R^{3d/(4-d)} \sqrt{\frac{N''}{2}} b^{y_{\rm h}}
  \psi''_{{\bf k}''={\bf 0}} \;.
\end{eqnarray}
We have introduced the coefficients $\tilde{r}_0$ and $\tilde{u}$, which denote
the location of the point $(r_0,u)$ in the new coordinates with respect to the
nontrivial (Ising) fixed point $\mu^*$ which we are now approaching (see
Fig.~\ref{fig:traject}).

The singular part of the free energy density, $f_{\rm s}$, is after the
transformation $\phi \to \psi$ denoted by $\tilde{f}_{\rm s}$,
\begin{equation}
  f_{\rm s}(r_0,u,h) =
  \tilde{f}_{\rm s}\left(\frac{r_0}{R^2}, \frac{u}{R^4}, \frac{h}{R}\right) \;.
\end{equation}
Furthermore, we introduce the quantity $\hat{f}_{\rm
  s}(\tilde{r}_0,\tilde{u},h) \equiv \tilde{f}_{\rm s}(r_0,u,h)$.  Because the
total free energy is conserved along the renormalization trajectory, the
singular part of the free energy density changes as
\begin{eqnarray}
\tilde{f}_{\rm s}\left( \frac{r_0}{R^2}, \frac{u}{R^4}, \frac{h}{R} \right)
  &=& l^{-d}\tilde{f}_{\rm s}\left(\frac{r_0}{R^2}l^2,
  \frac{u}{R^4}l^{4-d}, \frac{h}{R}l^{1+d/2}\right) \nonumber \\
  &=&  R^{-4d/(4-d)} \hat{f}_{\rm s} \left(\tilde{r}_0 R^{2d/(4-d)}, \tilde{u},
  h R^{3d/(4-d)} \right) \nonumber \\
  &=& b^{-d} R^{-4d/(4-d)} \hat{f}_{\rm s} \left(tR^{2d/(4-d)} b^{y_{\rm t}},
  \tilde{u}b^{y_{\rm i}}, hR^{3d/(4-d)} b^{y_{\rm h}}\right) \;,
\label{eq:enerscal}
\end{eqnarray}
where we have used the notation $t \equiv [T-T_{\rm c}(R)]/T_{\rm c}(R)$ for
$\tilde{r}_0$. In Fig.~\ref{fig:traject}, $t$ stands for the distance to the
critical line connecting $\mu_0^*$ and $\mu^*$.  In the second equality we have
substituted the crossover scale, $l=R^{4/(4-d)}$. Of course, this is only a
{\em qualitative\/} measure for the location of the crossover, but the
renormalization predictions for the scaling exponents are exact.  The
relation~(\ref{eq:enerscal}), which holds for $1<d<4$, is the key to the
scaling relations obtained on phenomenological grounds in Ref.~\cite{monbin}.
We will first illustrate this by deriving the $R$ dependence of the critical
amplitudes of the magnetization density $m$ and the magnetic susceptibility
$\chi$.  The magnetization density can be calculated by taking the first
derivative of the free energy density with respect to the magnetic scaling
field $h$,
\begin{eqnarray}
  m &=& \frac{\partial f_{\rm s}}{\partial h}(r_0,u,h)
     =  \frac{\partial \tilde{f}_{\rm s}}{\partial h}
        \left(\frac{r_0}{R^2},\frac{u}{R^4},\frac{h}{R}\right) \nonumber \\
    &=& b^{y_{\rm h}-d} R^{-d/(4-d)}
        \hat{f}^{(1)}_{\rm s}\left(tR^{2d/(4-d)} b^{y_{\rm t}},
        \tilde{u}b^{y_{\rm i}}, hR^{3d/(4-d)} b^{y_{\rm h}}\right) \;,
\label{eq:magn}
\end{eqnarray}
where $\hat{f}^{(1)}_{\rm s}$ denotes the first derivative of $\hat{f}_{\rm
  s}$ with respect to its third argument. Here we have neglected any nonlinear
relation between the magnetic scaling field $h$ and the physical magnetic
field. Furthermore, we have omitted a contribution from the analytic part of
the free energy, because $h$ only couples to the ${\bf k}={\bf 0}$ (uniform)
mode.  To extract the dependence of $m$ on $t$ and $R$ from
Eq.~(\ref{eq:magn}), we choose the rescaling factor $b$ such that the first
argument of the derivative of $\hat{f}_{\rm s}$ is equal to 1, i.e.,
$b=t^{-1/y_{\rm t}} R^{-2d/[y_{\rm t}(4-d)]}$, and set the irrelevant variable
$\tilde{u}$ and the magnetic scaling field $h$ equal to zero,
\begin{equation}
  m = t^\beta R^{(2d\beta-d)/(4-d)} \hat{f}^{(1)}_{\rm s}(1,0,0) \;.
\label{eq:magscal}
\end{equation}
This result agrees with Ref.~\cite[Eq.~(34)]{monbin}. In the same way we can
calculate the magnetic susceptibility from $\hat{f}_{\rm s}$ by taking the
second derivative with respect to $h$,
\begin{eqnarray}
\chi &=& \frac{\partial^2 f_{\rm s}}{\partial h^2}(r_0,u,h)
  = \frac{\partial^2 \tilde{f}_{\rm s}}{\partial h^2}
    \left(\frac{r_0}{R^2}, \frac{u}{R^4}, \frac{h}{R}\right) \nonumber \\
  &=& b^{2y_{\rm h}-d} R^{2d/(4-d)} \hat{f}^{(2)}_{\rm s}
    \left(tR^{2d/(4-d)} b^{y_{\rm t}}, \tilde{u}b^{y_{\rm i}},
    hR^{3d/(4-d)} b^{y_{\rm h}}\right) \;.
\label{eq:chi}
\end{eqnarray}
Choosing the arguments of the second derivative of $\hat{f}_{\rm s}$ as in
Eq.~(\ref{eq:magscal}), we find
\begin{equation}
  \chi = t^{-\gamma} R^{2d(1-\gamma)/(4-d)} \hat{f}^{(2)}_{\rm s}(1,0,0) \;,
\label{eq:chiscal}
\end{equation}
in agreement with Ref.~\cite[Eq.~(39)]{monbin}.  In Eqs.~(\ref{eq:magscal})
and~(\ref{eq:chiscal}) we have used the well-known relations between the
renormalization exponents and the critical exponents (see, e.g, Table~18 in
Ref.~\cite{ic3d}).

The finite-size scaling behavior of thermodynamic functions can also be derived
from this renormalization scenario by including a finite-size field $1/L$ as an
additional argument of the free energy density in Eq.~(\ref{eq:enerscal}).
Under the first renormalization transformation this scaling field will scale as
$l/L$ and under the second renormalization transformation as
$lb/L=bR^{4/(4-d)}L^{-1}$. The finite-size scaling behavior is found by
choosing $b$ such that $lb/L=1$, i.e., $b=LR^{-4/(4-d)}$. Substituting this
into Eqs.~(\ref{eq:magn}) and~(\ref{eq:chi}), we obtain for $m$
\begin{equation}
  m = L^{y_{\rm h}-d} R^{(3d-4y_{\rm h})/(4-d)}
  \hat{f}^{(1)}_{\rm s}
  \left( tL^{y_{\rm t}} R^{-2(2y_{\rm t}-d)/(4-d)},
  \tilde{u}L^{y_{\rm i}} R^{-4y_{\rm i}/(4-d)},
  hL^{y_{\rm h}} R^{(3d-4y_{\rm h})/(4-d)} \right) \;,
\label{eq:magfss}
\end{equation}
and for $\chi$,
\begin{equation}
  \chi = L^{2y_{\rm h}-d} R^{2(3d-4y_{\rm h})/(4-d)}
  \hat{f}^{(2)}_{\rm s}
  \left( tL^{y_{\rm t}} R^{-2(2y_{\rm t}-d)/(4-d)},
  \tilde{u}L^{y_{\rm i}} R^{-4y_{\rm i}/(4-d)},
  hL^{y_{\rm h}} R^{(3d-4y_{\rm h})/(4-d)} \right) \;.
\label{eq:chifss}
\end{equation}
These results agree with Ref.~\cite{monbin}, where the prefactors of the
magnetization density and the magnetic susceptibility were predicted as,
respectively, $L^{-\beta/\nu} R^{(2\beta-\gamma)/[\nu(4-d)]}$ and
$L^{\gamma/\nu} R^{(4\beta-2\gamma)/[\nu(4-d)]}$. Furthermore, the first
argument of the scaling functions was predicted as $tL^{1/\nu}R^{\cal K}$, with
${\cal K}=-(2\alpha)/[\nu(4-d)]$ (Ref.~\cite[Eq.~(25)]{monbin}). This is indeed
equivalent with our result ${\cal K}=-2(2y_{\rm t}-d)/(4-d)$.  However, the
predicted range dependence of the critical amplitudes (i.e., of the prefactors
in Eqs.~(\ref{eq:magfss}) and~(\ref{eq:chifss})) is only valid in the limit of
infinite range. For smaller ranges, $R$-dependent correction terms are present.
These correction terms can be calculated as well. They do {\em not\/} come from
the dependence of the scaling functions on the irrelevant fields, as
corrections to scaling normally do: these corrections vanish in the
thermodynamic limit. However, they come from higher-order contributions to the
renormalization of the $\psi^4$ coefficient which were previously neglected in
the derivation of the crossover scale $l_0$.  Note that in the neighborhood of
the Gaussian fixed point, the terms $\psi^n$ with $n< 2d/(d-2)$ are relevant
and that for $d = 2$ {\em all\/} higher-order terms are equally relevant.
However, the coefficients of these terms are, after the rescaling $\phi \to
\psi$, proportional to $R^{-n}$, so the leading contribution comes from the
term $wR^{-6}\psi^6$.  Under a spatial rescaling with a factor $l = e^s$, the
renormalization equation for this term is, to leading order,
\begin{equation}
\frac{dw'}{ds} = (6-2d) w' \;.
\end{equation}
The solution of this equation, $w'(s)=w e^{(6-2d)s}$ can be substituted in the
renormalization equation for the $\psi^4$ coefficient,
\begin{equation}
\frac{1}{R^4} \frac{du'}{ds} = (4-d) \frac{u'}{R^4} + a \frac{w'}{R^6} \;.
\label{eq:rg-u}
\end{equation}
To first order in $w$, this yields,
\begin{equation}
\frac{u'}{R^4} =
 e^{(4-d)s} \frac{1}{R^4}
 \left[ u + \frac{a}{2-d} \frac{w}{R^2} \left( e^{(2-d)s}-1 \right) \right]
 =  l^{4-d} \frac{1}{R^4}
 \left[ u + \frac{a}{2-d} \frac{w}{R^2} \left( l^{2-d}-1 \right) \right] \;,
\label{eq:leadcorr}
\end{equation}
where $u$ and $w$ denote the values of $u'$ and $w'$ at $l=1$, respectively.
This implies that the previously obtained crossover scale $l_0=R^{4/(4-d)}$
is multiplied by a factor $(1+ \tilde{a}R^{-2})$ and hence all critical
amplitudes will exhibit this correction.
However, the solution~(\ref{eq:leadcorr}) is not valid for $d=2$, where
$uR^{-4}\psi^4$ and $wR^{-6}\psi^6$ are equally relevant. The solution
of Eq.~(\ref{eq:rg-u}) is then given by
\begin{equation}
\frac{u'}{R^4} = e^{2s} \frac{1}{R^4} \left( u + a \frac{w}{R^2} s \right)
 = l^2 \frac{1}{R^4} \left( u + a \frac{w}{R^2} \ln l \right) \;,
\label{eq:2dcorr}
\end{equation}
which yields a (leading) correction factor $[1+ R^{-2} (\tilde{a}_1 +
\tilde{a}_2 \ln R)]$ in the crossover scale and the critical amplitudes.

{}From a similar mechanism we can derive the $R$ dependence of the so-called
shift of the critical temperature~\cite[Eq.~(15)]{monbin}.  A detailed
treatment of the shift of $T_{\rm c}$ can be found in, e.g.,
Refs.~\cite{bzj,univers}. It arises from the $u$-dependent contribution in the
renormalization equation for the $\psi^2$ term,
\begin{equation}
 \frac{1}{R^2} \frac{dr_0'}{ds} = 2\frac{r_0'}{R^2} + c \frac{u'}{R^4} \;.
\label{eq:rg-r0}
\end{equation}
Thus, the first argument on the RHS of the first equation
in~(\ref{eq:enerscal}) should be replaced by
\begin{equation}
  \frac{r_0'}{R^2} =
  l^2 \left[ \left( \frac{r_0}{R^2} - \tilde{c}\frac{u_0}{R^4} \right)
    + \tilde{c}\frac{u_0}{R^4} l^{2-d} \right]
  = l^2 \frac{1}{R^2} \left[ \left( r_0 - \tilde{c}\frac{u_0}{R^2} \right)
    + \tilde{c} \frac{u_0}{R^2} l^{2-d} \right]  \;.
\label{eq:tc-shift}
\end{equation}
The term between round brackets is proportional to the reduced temperature and
the last term is the leading shift.  Substitution of the crossover scale $l_0$
shows that this shift in the reduced temperature is proportional to
$R^{-2d/(4-d)}$, which indeed vanishes in the mean-field ($R \to \infty$)
limit.  Remarkably, this disagrees with Ref.~\cite{monbin}, where a shift
$\propto R^{-d}$ was predicted. Unfortunately, it is not possible to settle
this issue at present by means of Monte Carlo data, because only results for
$d=2$ are available, which is a special case.  Viz., for $d=2$ we obtain
instead of Eq.~(\ref{eq:tc-shift}) the following solution of
Eq.~(\ref{eq:rg-r0}),
\begin{equation}
  \frac{r_0'}{R^2} =
  l^2 \left( \frac{r_0}{R^2} + c \frac{u_0}{R^4} \ln l \right)
  = l^2 \frac{1}{R^2} \left( r_0 + c \frac{u_0}{R^2} \ln l \right) \;.
\end{equation}
Thus, we find, upon substitution of the crossover scale, that the shift in
the reduced temperature has the form $(p + q \ln R)/R^2$, where the constant
$p$ comes from a multiplicative factor introduced by the crossover criterion.
In Ref.~\cite{monbin}, $d=2$ was already suggested as a special case, with
possibly logarithmic corrections. The renormalization argument shows that such
a $\ln R$ term is indeed present.

Now, let us return to Eq.~(\ref{eq:trunc}), where we omitted quartic (and
higher) terms in $kR$. It follows from the renormalization scenario that terms
proportional to $k^{2n}$ transform as $k^{2n}l^{2-2n}$ under the first
renormalization transformation and hence are irrelevant for $n>1$.
The behavior of these terms under the second renormalization transformation is
less simple, but again quartic and higher terms do not influence the leading
terms, see, e.g., Ref.~\cite[Section~VII.6]{ma}.

Besides, it can be seen that the structure of the interaction term does not
depend on the details of the spin--spin interaction.  E.g., replacing the
interaction term~(\ref{eq:cstint}) with $K({\bf r}) = cR^{-d}\exp \left[
  -(r/R)^2 \right]$ leads to precisely the same structure of the LGW
Hamiltonian and hence to the same scaling relations involving $R$. This is in
agreement with the universality hypothesis.

Furthermore, the renormalization description explains why the interaction range
$R$ must be large to observe the predicted powers of $R$ in the critical
amplitudes: only for systems with $R$ large the renormalization trajectory
starts in the neighborhood of the Gaussian fixed point and hence only these
systems will accurately display the corresponding $R$ dependence. Finally, in
the finite-size scaling description, the system size must be sufficiently large
in order to observe the crossover to Ising-like critical behavior: we require
that the rescaling factor $b$ is minimal of order unity, or $L = {\cal O}
(R^{4/(4-d)})$.

\section{Monte Carlo results and comparison with the theoretical predictions}
\label{sec:mc}

\subsection{Definition of the model}
We have carried out Monte Carlo simulations for two-dimensional Ising systems
consisting of $L \times L$ lattice sites with periodic boundary conditions and
an extended range of interaction.  Each spin interacts equally with its $z$
neighbors lying within a distance $R_m$, as defined in Eqs.~(\ref{eq:latham})
and~(\ref{eq:cstint}) with $R$ replaced by $R_m$ and $d=2$.  The Monte Carlo
simulations were carried out using a special cluster algorithm for long-range
interactions~\cite{ijmpc}.  Its application to the interactions defined above
is described in Appendix~\ref{sec:algorithm}.  Following Ref.~\cite{monbin} we
define the effective range of interaction $R$ as
\begin{eqnarray}
\label{eq:effrange}
R^2 &\equiv& \frac{\sum_{j \neq i} ({\bf r}_i - {\bf r}_j)^2 K_{ij}}%
                  {\sum_{j \neq i} K_{ij}} \\
    &=& \frac{1}{z} \sum_{j \neq i} |{\bf r}_i - {\bf r}_j |^2
        \quad\quad \text{with } |{\bf r}_i - {\bf r}_j | \leq R_m \;.
\end{eqnarray}
Table~\ref{tab:range} lists the values of $R_m^2$ for which we have carried out
simulations, as well as the corresponding values of $R^2$.  The ratio between
$R^2$ and $R_m^2$ approaches $1/2$, as can be simply found when the sums in
Eq.~(\ref{eq:effrange}) are replaced by integrals.  Note that the results for
$R_m^2=18$ and $R_m^2=32$ cannot be compared to those presented by Mon and
Binder, because in Ref.~\cite{monbin} the interactions were for these two
system sizes spatially distributed within a {\em square}, as can be seen from
the number of interacting neighbors and the corresponding effective ranges of
interaction.

\subsection{Determination of the critical temperature}
The critical temperatures $T_{\rm c}$ of these systems have been determined
using the well-known universal amplitude ratio $Q_L \equiv \langle m^2
\rangle^2_L / \langle m^4 \rangle_L$. Both in the Ising and in the mean-field
limit the critical-point value of this quantity is accurately known.  In the
mean-field limit, $Q=Q_{\rm MF} = 8\pi^2 / [\Gamma(\frac{1}{4})]^4 \approx
0.456947$, see Refs.~\cite{bzj} and~\cite[Appendix A]{ijmpc}. In the Ising
limit, $Q = Q_{\rm I} = 0.856216~(1)$~\cite{kamblo}.

\begin{figure}
\begin{center}
\leavevmode
\epsfbox{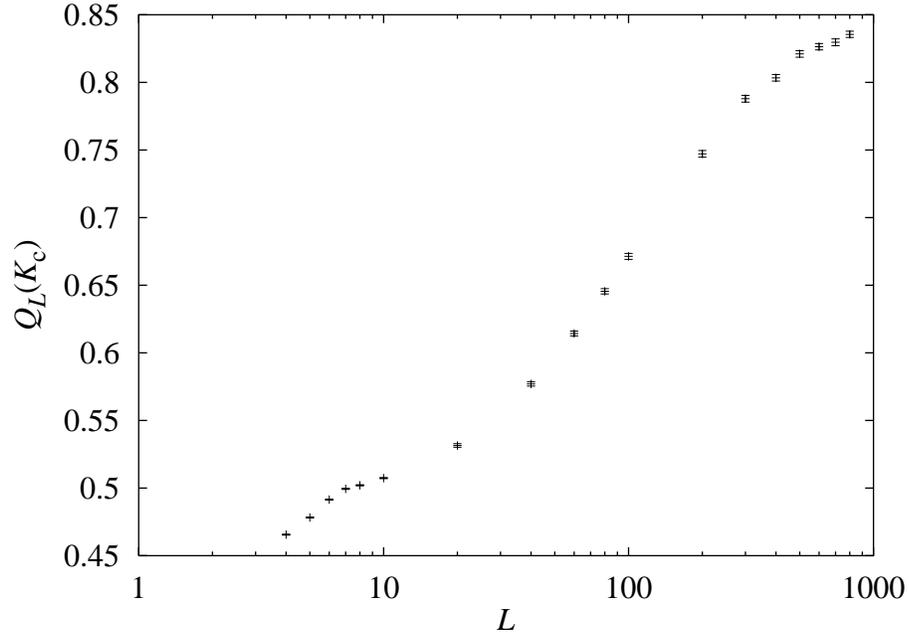} 
\end{center}
\caption{The critical-point amplitude ratio $Q_L(K_{\rm c})$ for $R_m^2=140$ as
  function of the system size. For large $L$, $Q_L(K_{\rm c})$ approaches the
  Ising limit $Q_{\rm I} \approx 0.856216$.  For small $L$, $Q_L(K_{\rm c})$
  approaches the mean-field limit $Q_{\rm MF} \approx 0.456947$, although there
  are significant finite-size corrections for the smallest values of $L$.}
\label{fig:cross-q}
\end{figure}

As was noted in Section~\ref{sec:theory} and also in Ref.~\cite{monbin}, rather
large system sizes (${\cal O}(R^2)$) are required to determine $T_{\rm c}$,
since $Q$ must approach $Q_{\rm I}$. For $R_m^2 \leq 10$ we have included
linear system sizes up to $L=500$ and for larger ranges we have used system
sizes up to $L=700$ or even $L=800$ ($R_m^2=100, 140$).  For each run we have
generated $10^6$ Wolff clusters after equilibration of the system. The various
thermodynamic quantities were sampled after every 10th Wolff cluster.  In
Fig.~\ref{fig:cross-q}, $Q_L(K_{\rm c})$ for $R_m^2=140$ is plotted as function
of the system size. One clearly observes the crossover from $Q_{\rm MF}$ (for
$L \ll R_m^2$) to $Q_{\rm I}$ (for $L \gg R_m^2$).

\begin{figure}
\begin{center}
\leavevmode
\epsfbox{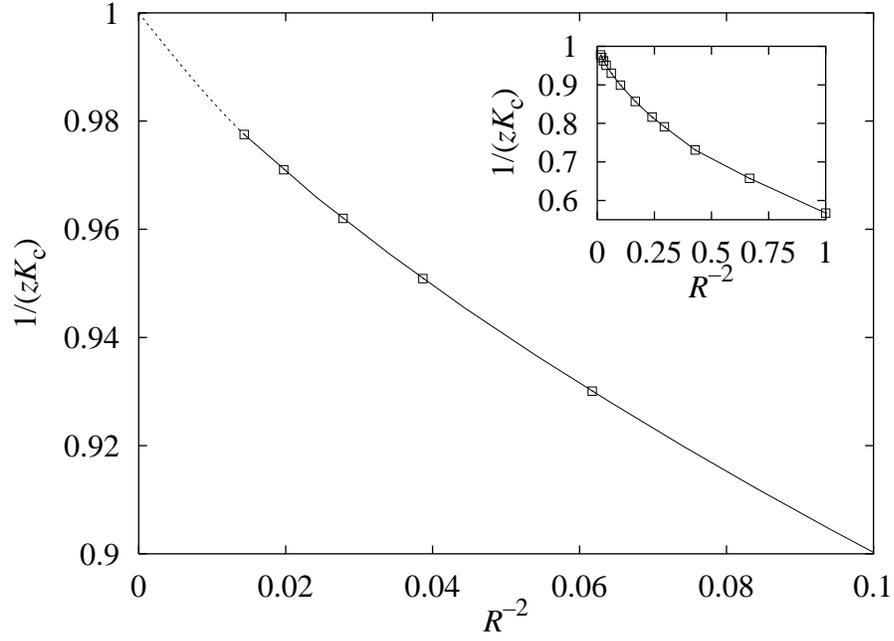} 
\end{center}
\caption{Plot of $1/(zK_{\rm c})$ versus $R^{-2}$. The dashed line denotes the
  extrapolation to the mean-field limit. The inset shows $1/(zK_{\rm c})$ over
  the full range of $R^{-2}$ between the Ising and the mean-field limit.}
\label{fig:kc-r2}
\end{figure}

\begin{figure}
\begin{center}
\leavevmode
\epsfbox{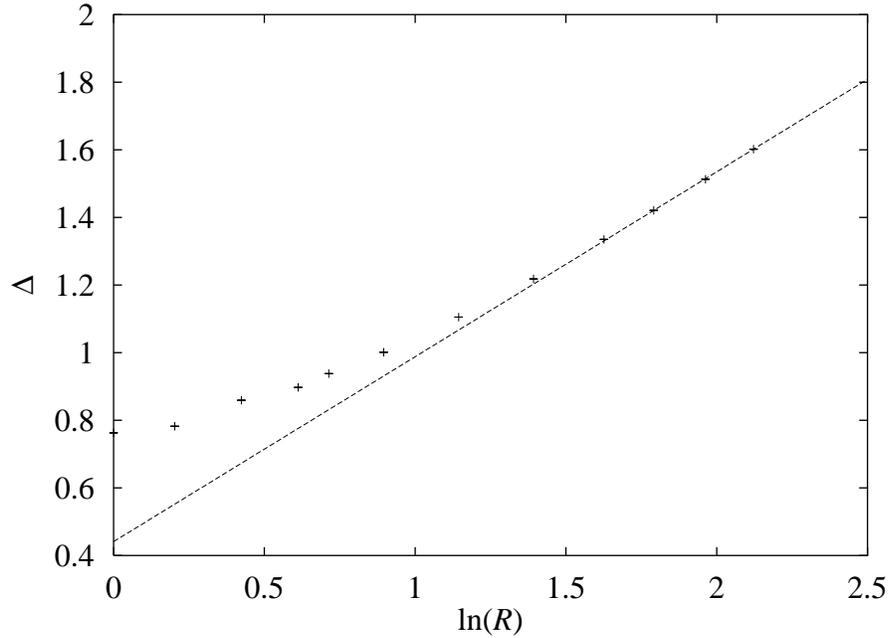} 
\end{center}
\caption{$\Delta \equiv (zK_{\rm c}-1)R^2$ versus $\ln R$. For large $R$ the
  graph strongly suggests the presence of a logarithmic correction in the shift
  of the critical temperature. The error bars do not exceed the symbol size.}
\label{fig:logshift}
\end{figure}

In the Ising limit, the finite-size expansion of $Q_L$ reads (see, e.g.,
Ref.~\cite{ic3d})
\begin{equation}
  Q_L(K)= Q + a_1 (K-K_{\rm c}) L^{y_{\rm t}}
        + a_2 (K-K_{\rm c})^2 L^{2y_{\rm t}} + \cdots
        + b_1 L^{y_{\rm i}} + b_2 L^{y_2} + \cdots \;.
\label{eq:expan}
\end{equation}
$K$ denotes the spin--spin coupling, $K_{\rm c}$ the critical coupling, and the
$a_i$ and $b_i$ are non-universal (range-dependent) coefficients. The term
proportional to $L^{y_2}$, with $y_2=d-2y_{\rm h}$, comes from the field
dependence of the analytic part of the free energy. In a $\phi^4$ theory this
term is absent, as was stated in section~\ref{sec:theory}, but in a discrete
model it should be allowed for. The exponents $y_{\rm t}$, $y_{\rm h}$, and
$y_{\rm i}$, which have already been introduced in the previous section, are,
respectively, the temperature, magnetic, and leading irrelevant exponent for
the two-dimensional Ising model; $y_{\rm t}=1$, $y_{\rm h}=\frac{15}{8}$, and
$y_{\rm i}=-2$. Table~\ref{tab:qfit} displays the results of a least-squares
fit according to Eq.~(\ref{eq:expan}), where $y_{\rm t}$, $y_{\rm i}$, and
$y_{\rm 2}$ were kept fixed at their theoretical values.  For comparison we
have included the estimates for $K_{\rm c}$ from Ref.~\cite{monbin}. Except for
$R_m^2=10$ there is good agreement between the respective estimates.  The
discrepancy for $R_m^2=10$ may be explained by the limited range of
system sizes in Ref.~\cite{monbin}.  Furthermore, for $R_m^2=2$, which
corresponds to the Ising model with nearest and next-nearest neighbor
interactions, an accurate transfer-matrix estimate of the critical coupling
exists, $K_{\rm c} = 0.19019269(5)$~\cite{nighblo}. The Monte Carlo result
agrees with this value.  The results for $Q$ are in good agreement with the
expected value $Q_{\rm I}$, which confirms not only that universality is
satisfied, but also that the maximum system sizes in our simulations are
sufficiently large, so that crossover to Ising-like critical behavior indeed
has taken place, as it should for an accurate determination of $K_{\rm c}$. In
fact, the error margins on $K_{\rm c}$ can be reduced significantly by fixing
$Q$ at its Ising value in Eq.~(\ref{eq:expan}) (see Table~\ref{tab:qfit}).
Figure~\ref{fig:kc-r2} illustrates the shift in $1/(zK_{\rm c}) \propto T_{\rm
  c}$ as function of $R^{-2}$.  Even close the mean-field limit ($R^{-2} \to
0$), the deviation of $1/(zK_{\rm c})$ from $1$ seems not truly linear.
Therefore we have tried to identify the logarithmic term, which was suggested
in Ref.~\cite{monbin} and derived from the renormalization scenario, by writing
the following expression for the critical coupling,
\begin{equation}
zK_{\rm c} = 1 + \frac{p + q \ln R}{R^2} \;.
\end{equation}
In Fig.~\ref{fig:logshift} we have plotted $\Delta \equiv (zK_{\rm c}-1) R^2$
versus $\ln R$. Indeed, for large values of $R$ the points lie approximately on
a straight line, confirming the presence of the logarithmic
correction.

\begin{figure}
\begin{center}
\leavevmode
\epsfbox{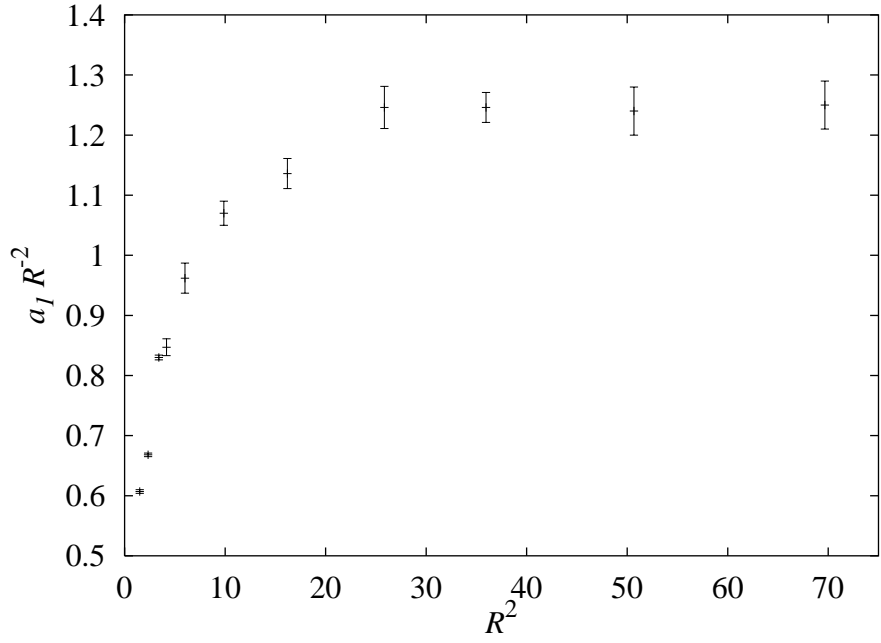} 
\end{center}
\caption{Range dependence of the amplitude of the temperature-dependent
  argument of the finite-size scaling function of the universal amplitude ratio
  $Q$.}
\label{fig:coef}
\end{figure}

Another $\ln R$ correction was suspected in Ref.~\cite{monbin} in the
temperature-dependent argument of the finite-size scaling functions. This
argument is proportional to $R^{-2(2y_{\rm t}-d)/(4-d)} =
R^{-2\alpha/[\nu(4-d)]}$, see Eqs.~(\ref{eq:magfss}) and~(\ref{eq:chifss}). For
$d=2$, $\alpha=0$ implies a logarithmic divergence of the specific heat and
hence one might expect a similar logarithmic factor here. On the other hand, we
have not found a mechanism in the renormalization scenario which could explain
such a factor. Therefore, we have numerically examined the range dependence of
the coefficient $a_1$ in Eq.~(\ref{eq:expan}). Since $(K-K_{\rm c})$ is
proportional to $R^{-2}$, we must first divide $a_1$ by $R^2$.
Figure~\ref{fig:coef} displays this quantity as function of the range. For
small ranges, there is a strong dependence on $R$, but the coefficients seem to
approach a constant value in the large-range limit. This suggests that a
logarithmic correction factor is absent.

\subsection{Range dependence of the magnetization density}
\begin{figure}
\begin{center}
\leavevmode
\epsfbox{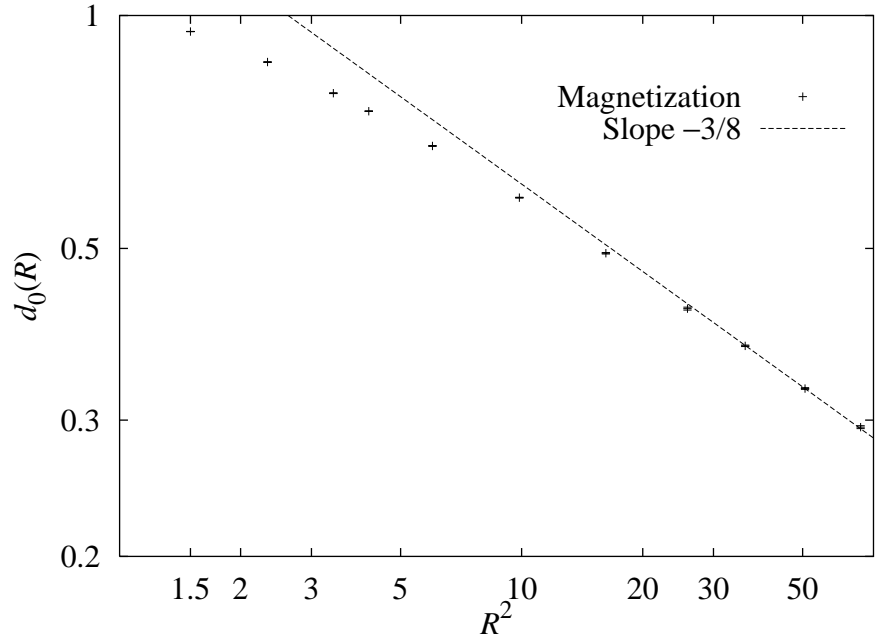} 
\end{center}
\caption{The critical amplitude $d_0(R)$ of the magnetization density versus
  $R^2$.}
\label{fig:mag}
\end{figure}

We have sampled the absolute magnetization density, $\langle |m| \rangle$, for
which the range dependence is given by Eq.~(\ref{eq:magfss}). This quantity
has been fitted to the following finite-size expansion,
\begin{equation}
 m_L(K,R)= L^{y_{\rm h}-d} \left\{ d_0(R) + d_1(R) [K-K_{\rm c}(R)] L^{y_{\rm
        t}} + d_2(R) [K-K_{\rm c}(R)]^2 L^{2y_{\rm t}} + \cdots
        + e_1(R) L^{y_{\rm i}} + \cdots \right\} \;,
\label{eq:mag-expan}
\end{equation}
where we now have explicitly indicated the range dependence of the parameters.
The critical couplings found from this quantity agree well with those obtained
from the amplitude ratio $Q$ and the exponent $y_{\rm h}$, listed in
Table~\ref{tab:mag}, is in good agreement with the exact value $15/8$.
Furthermore, we have made a least-squares fit with $K_{\rm c}$ fixed at the
most accurate values obtained from $Q$. The corresponding estimates for $y_{\rm
  h}$ are also shown in Table~\ref{tab:mag}. They lie even closer to $15/8$,
which again corroborates that all systems belong to the Ising universality
class. From the critical amplitudes $d_0(R)$ we can derive the leading $R$
dependence of the magnetization. To increase the accuracy, the values in
Table~\ref{tab:mag} were determined with $y_{\rm h}$ fixed at its theoretical
value. As can be seen from the log--log plot in Fig.~\ref{fig:mag}, the
approach to the asymptotic scaling behavior is very slow. Therefore we have
determined the scaling exponent in two different ways. A straight line through
the points for the three largest ranges yielded $d_0(R) \propto R^{-0.738
  (13)}$, in agreement with the predicted exponent $-3/4$
(Eq.~(\ref{eq:magfss})). Inclusion of the correction factor $[ 1+ R^{-2}
(\tilde{a}_1 + \tilde{a}_2 \ln R)]$, as predicted from Eq.~(\ref{eq:2dcorr}),
allowed us to include {\em all\/} data points in the fit and yielded $d_0(R)
\propto R^{-0.756 (5)}$, also in good agreement with the predicted exponent.

\subsection{Range dependence of the susceptibility}
\begin{figure}
\begin{center}
\leavevmode
\epsfbox{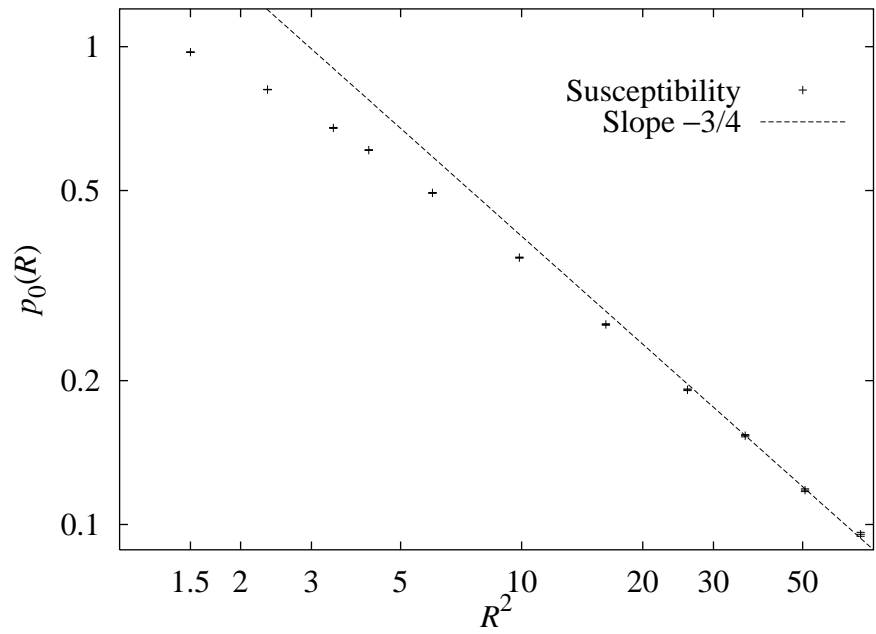} 
\end{center}
\caption{The critical amplitude $p_0(R)$ of the magnetic suscetibility versus
  $R^2$.}
\label{fig:chi}
\end{figure}

The magnetic susceptibility can be calculated from the average square
magnetization,
\begin{equation}
  \chi = L^d \langle m^2 \rangle \;.
\end{equation}
We thus expect the following finite-size scaling behavior,
\begin{equation}
 \chi_L(K,R)= s_0 + L^{2y_{\rm h}-d} \left\{ p_0(R) + p_1(R) [K-K_{\rm c}(R)]
              L^{y_{\rm t}} + p_2(R) [K-K_{\rm c}(R)]^2 L^{2y_{\rm t}} + \cdots
              + q_1(R) L^{y_{\rm i}} + \cdots \right\} \;.
\label{eq:chi-expan}
\end{equation}
The term $s_0$ comes from the analytic part of the free energy.  Because it
tends to interfere with the term proportional to $q_1(R)$, we have ignored it
in the further analysis.  Again, the critical couplings obtained from a
least-squares fit lie close to those in Table~\ref{tab:qfit} and the estimates
for $y_{\rm h}$ agree with the Ising value (see Table~\ref{tab:chi}).  By
repeating the fits with $K_{\rm c}$ fixed at the most accurately known values,
the values for $y_{\rm h}$ lie even closer to $15/8$ (third column of
Table~\ref{tab:chi}). From the parameter $p_0(R)$, plotted in
Fig.~\ref{fig:chi}, we can extract the leading range dependence of the
susceptibility.  A straight line through the amplitudes for the three largest
ranges gave $p_0(R) \propto R^{-1.46(3)}$.  For a curve (including the first
correction term) through the amplitudes it was necessary to include the data
for all ranges $R^2 \geq 7/3$ in the fit, in order to determine the coefficient
of the $\ln R$ factor. This yielded $p_0(R) \propto R^{-1.47 (2)}$.  Both
exponents are in good agreement with the predicted value $2(3d-4y_{\rm
  h})/(4-d) = -3/2$.

\subsection{Spin--spin correlation function}
\begin{figure}
\begin{center}
\leavevmode
\epsfbox{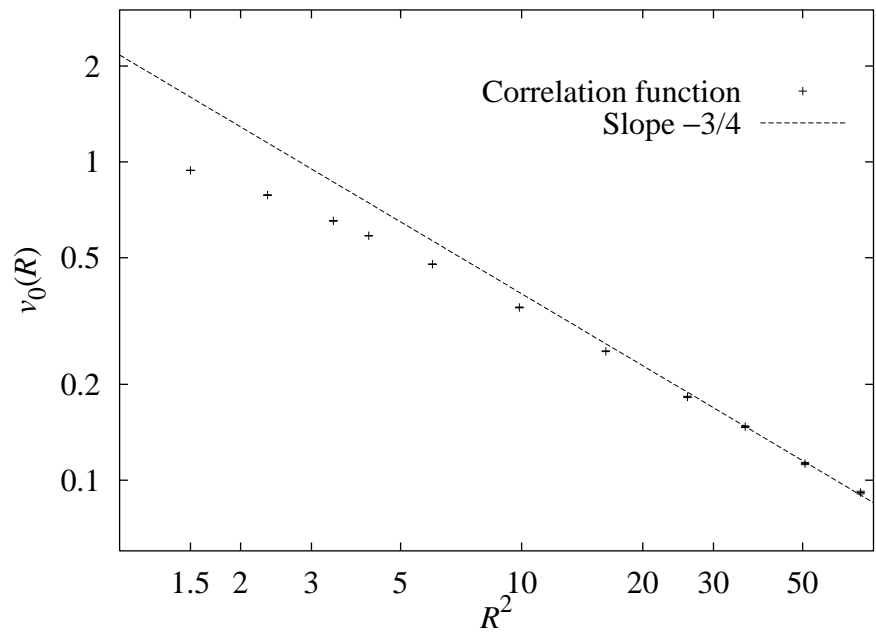} 
\end{center}
\caption{The critical amplitude $v_0(R)$ of the spin--spin correlation function
  versus $R^2$.}
\label{fig:corr}
\end{figure}

The finite-size scaling behavior of the spin--spin correlation function $g({\bf
  r})$ closely resembles that of the magnetic susceptibility $\chi$, as may be
expected from the fact that $\chi$ is the spatial integral of $g$. We also
expect the range dependence of the two quantities to be the same.
We have sampled the correlation function over half the system size and analyzed
it using the expansion
\begin{equation}
 g_L(K,R)= L^{2y_{\rm h}-2d} \left\{ v_0(R) + v_1(R) [K-K_{\rm c}(R)]
              L^{y_{\rm t}} + v_2(R) [K-K_{\rm c}(R)]^2 L^{2y_{\rm t}} + \cdots
              + w_1(R) L^{y_{\rm i}} + \cdots \right\} \;.
\end{equation}
The constant term in~(\ref{eq:chi-expan}) is not present here (see, e.g,
Ref.~\cite{ic3d}). Table~\ref{tab:corr} shows the results for $y_{\rm h}$, both
with $K_{\rm c}$ free and fixed. In the latter case, $y_{\rm h}$ is in accurate
agreement with its theoretical value, just as for the magnetization density and
the magnetic susceptibility. Figure~\ref{fig:corr} shows a log--log plot of the
critical amplitude $v_0(R)$ as function of the range. A fit of a straight line
through the points with $R^2 > 35$ (i.e., $R_m^2 \geq 72$) yielded $v_0(R)
\propto R^{-1.46 (3)}$, whereas a curve through all points with $R^2 > 7/3$
gave $v_0(R) \propto R^{-1.49 (2)}$. Both estimates are again in good agreement
with the predicted exponent $-3/2$.

\section{Conclusion}
\label{sec:concl}
In this paper, we have derived the dependence of scaling functions on the range
of interactions from renormalization-group arguments. The results agree with
the predictions of Mon and Binder and yield in addition the corrections to the
leading scaling behavior, including the previously conjectured logarithmic
factor in the shift of the critical temperature of two-dimensional systems.

We have also carried out accurate Monte Carlo simulations for systems in which
the range of the interactions was large enough to verify the theoretical
predictions.  It was confirmed with high precision that all examined systems
belong to the 2D Ising universality class.  Besides the range dependence of
critical amplitudes, we also observed the predicted range dependence of the
corrections to scaling.

\appendix
\section{Fourier transform of a spherically shaped interaction profile}
\label{sec:fourier}
We define the following isotropic spin--spin interaction $K_d$ (the
subscript $d$ denotes the dimensionality),
\begin{equation}
K_d({\bf r}) \equiv \left\{ \begin{array}{ll}
                              cR^{-d} & \mbox{if $|{\bf r}| \leq R$} \\
                              0 & \mbox{if $|{\bf r}| > R$}
                            \end{array}
                    \right.
\label{eq:cstint}
\end{equation}
We have normalized the interaction strength, to make the integrated interaction
(energy) independent of the range. In this appendix, we calculate the Fourier
transform of this interaction for a general number of dimensions.  For $d=1$
the calculation is trivial:
\begin{equation}
\tilde{K}_1(k) = \frac{c}{R} \int_{-R}^{+R} dx \; e^{ikx} =
  \frac{2c}{kR} \sin(kR) \;.
\label{eq:block-1}
\end{equation}
For $d=2$ and $d=3$ one obtains Bessel functions. Using the equality
$J_{1/2}(x)= \sqrt{2/(\pi x)} \sin(x)$, the results for $d=1,2,3$ can be
summarized as
\begin{equation}
  \tilde{K}_d({\bf k}) = c \left(\frac{2\pi}{kR}\right)^{d/2} J_{d/2} (kR) \;,
\label{eq:block-d}
\end{equation}
where $J_\nu$ is a Bessel function of the first kind of order $\nu$.
This suggests that this equality is valid for general $d$, which can indeed be
shown by induction. If we assign the $x$ coordinate to the $(d+1)$th spatial
dimension and use the notation $k_d^2 = \sum_{j=1}^{d} k_i^2$, we may write
\begin{eqnarray}
  \tilde{K}_{d+1}({\bf k}) &=& \frac{c}{R^{d+1}} \int_{-R}^{+R} dx
    \cos(k_x x) \left(\frac{2\pi}{k_d}\right)^{d/2} (R^2-x^2)^{d/4}
    J_{d/2} \left(k_d \sqrt{R^2-x^2}\right) \nonumber \\
  &=& \frac{2c}{R^{d+1}} \left(\frac{2\pi}{k_d}\right)^{d/2}
    \int_0^{R} dp \cos\left(k_x \sqrt{R^2-p^2}\right)
    \frac{p^{(d+2)/2}}{\sqrt{R^2-p^2}} J_{d/2}(k_d p) \nonumber \\
  &=& c \left(\frac{2\pi}{kR}\right)^{(d+1)/2} J_{(d+1)/2}(kR) \;,
\end{eqnarray}
where we have used a Hankel transform of general order, see, e.g.,
Ref.~\cite[p.~40, Eq.~(48)]{erdelyi}.

\section{Monte Carlo algorithm for spin systems with medium-range interactions}
\label{sec:algorithm}
The cluster algorithm we have used for the present Monte Carlo simulations has
essentially been described in Ref.~\cite{ijmpc}.  However, that description is
rather concise; here, we present a somewhat more elaborate discussion of the
mathematical backgrounds, and we outline how the algorithm is applied to
medium-range models of the type studied in this paper.

The description is given in terms of the Wolff cluster algorithm~\cite{wolff},
but the principle applies only to the cluster formation process. Thus, it is
also applicable in the Swendsen--Wang case~\cite{SW}.  For simplicity, we
describe the way a cluster of spins is built in the case that there are only
ferromagnetic interactions.

For each spin in the cluster, we have to run a task described below. During
this task, new spins may be included in the cluster. For this reason, it is
convenient to use a ``stack'' memory containing the addresses of the spins for
which the task remains to be done.

The task for a spin $s_i$ read from the stack is the following. A loop is
executed over all neighbors $s_j$ interacting with $s_i$. In each step of this
loop, the bond connecting sites $i$ and $j$ is ``activated'' with a probability
\begin{equation}
 \label{eq:bondprob}
 p(s_i,s_j) = \delta_{s_i s_j} p \;,
\end{equation}
where $p \equiv 1-\exp(-2K)$, in which $K$ is the coupling between $s_i$ and
$s_j$.  The simulation process would conventionally include a test whether
$s_i$ and $s_j$ are parallel, and if so, the production of a uniformly
distributed pseudo-random number $r$. If $r<p$, the bond is activated, $s_j$ is
added to the cluster, and its address $j$ is stored in the stack memory.  Since
this loop runs over all neighbors interacting with $s_i$, i.e., over all sites
within a distance $R_m$, the process becomes very time consuming when the range
$R_m$ of the interactions becomes large, just as in the case of Metropolis
simulations.

However, the cluster formation process can be formulated in a more efficient
way. Part of the work involved in the activation of the bonds between $s_i$ and
its neighbors $s_j$ can be done in a way that does not depend on the signs of
the spins. Thus, as a first step, the bonds connected to $s_i$ are
``provisionally activated'' with a probability $p$, independent of their
relative sign.  Typically, only a small number of bonds will be provisionally
activated for each entry $i$ in the stack memory (i.e., each spin in the
cluster).  Then, in the second step, the provisionally active bonds, say
between sites $i$ and $j$, are actually activated if $s_i=s_j$, i.e., with
probability $\delta_{s_i s_j}$. During the second step, the bonds that were
left inactive in the first step can be ignored.

Since the first step does not depend on the signs of the spins, and the
probability $p$ is typically quite small, we introduce (following
Ref.~\cite{ijmpc}) a {\em cumulative bond probability}. This quantity
determines which bond is the {\em next\/} bond to be provisionally activated.
The probability that, during the first step, $k-1$ bonds are left inactive and
that the $k$th bond is provisionally activated is equal to
\begin{equation}
\label{eq:firstbond}
  P(k) = (1-p)^{k-1} p \;.
\label{eq:bprob}
\end{equation}
The cumulative bond probability $C(k)$ is defined as
\begin{equation}
\label{eq:cumprob}
  C(k) \equiv \sum_{n=1}^{k} P(n) \;.
\end{equation}
The interval $k-1$ to the next bond to be provisionally activated is obtained
by drawing a random number $r$ ($\in [0,1\rangle$). If this random number lies
between $C(k-1)$ and $C(k)$, $k-1$ bonds are skipped and the $k$th bond is
provisionally activated.  It is readily seen that this procedure leads to
precisely the required probabilities given in Eq.~(\ref{eq:bprob}). But the
number of operations per spin in the cluster is only of order $p R_m^2$; near
criticality, this quantity is approximately equal to 1. Thus, the work involved
in the decision concerning the actual bond activation is also of order unity.

We check independently whether the resulting probability of activating the
first bond at a distance $k$ is equal to that in the conventional approach.
Consider a cluster spin, say $s_0$, with a chain of neighbors denoted $s_1
\ldots s_{k}$, of which $m$ spins are antiparallel to $s_0$ and $l=k-m$ spins
are parallel to $s_0$, among which $s_k$. In the conventional Wolff cluster
algorithm, the probability that $s_k$ is the {\em first\/} spin to be added to
the cluster is, provided that $s_0=s_k$, given by
\begin{equation}
\label{eq:convtotprob}
 1^m (1-p)^{l-1} p = (1-p)^{l-1} p \;.
\end{equation}
On the other hand, if we use the cumulative bond probability
(\ref{eq:cumprob}), this probability is calculated as follows. Either the $k$th
spin is selected directly (if the first random number lies between $C(k-1)$ and
$C(k)$) or one of the $m$ antiparallel spins is selected, say $s_a$, which is
of course not added to the cluster. In the latter case, another random number
is drawn and a new spin is selected. Again, this may be the $k$th spin, or one
of the remaining antiparallel spins between $s_a$ and $s_k$.  Now, let us show
that the sum of these probabilities of adding $s_k$ as the first spin to the
cluster is equal to~(\ref{eq:convtotprob}).  Denote the number of selected,
``intermediate'', antiparallel spins by $i$.  There are
(\raisebox{-0.5ex}{$\stackrel{\scriptstyle m}{\scriptstyle i}$}) possibilities
of selecting $i$ intermediate spins. The probability of selecting $s_k$ after
each of these sequences of spins is $p^i (1-p)^{m-i} (1-p)^{l-1} p$.  The total
probability is the sum over all numbers of intermediate spins
\begin{equation}
\label{eq:clustotprob}
 \sum_{i=0}^{m} \left(
      \mbox{\raisebox{-0.95ex}{$\stackrel{\textstyle m}{\textstyle i}$}}
      \right) p^i (1-p)^{m-i} (1-p)^{l-1} p
      = (1-p)^{l-1} p \;,
\end{equation}
which is indeed equal to (\ref{eq:convtotprob}).

As shown in Ref.~\cite{ijmpc}, $C_j(k)=1 - \exp (-\sum_{n=j+1}^{k} K) = 1 -
\exp(-[k-j]K )$. By inverting this relation, the bond distance $k$ can be
calculated from $C_j(k)$, i.e., from the random number.  This approach is
highly efficient; compared to conventional (Metropolis) algorithms the gain is
typically a factor ${\cal O} (R_m^d L^2)$.  Finally we remark that efficient
variants of this technique can be applied to long-range O($n$) models with
$n>1$. Again, bonds are provisionally activated with a probability $p \equiv 1-
\exp(-2K)$; actual activation of a bond between spins $\vec{s}_i$ and
$\vec{s}_j$ is done afterwards with a probability $[1- \exp(-2K
s_{i,z}s_{j,z})]/p$ if $(\vec{s}_i \cdot \vec{s}_j) > 0$ (where $z$ defines the
spin-flip direction of the cluster-formation step) and otherwise with
probability $0$.

\newpage
\begin{table}
\renewcommand{\arraystretch}{1.2}
\begin{tabular}{r|r|c}
$R_m^2$  & $z$ & $R^2$               \\ \hline
 2       &   8 & $\frac{3}{2}$       \\
 4       &  12 & $\frac{7}{3}$       \\
 6       &  20 & $\frac{17}{5}$      \\
 8       &  24 & $\frac{25}{6}$      \\
10       &  36 & 6                   \\
18       &  60 & $\frac{148}{15}$    \\
32       & 100 & $\frac{81}{5}$      \\
50       & 160 & $\frac{517}{20}$    \\
72       & 224 & $\frac{1007}{28}$   \\
100      & 316 & $\frac{4003}{79}$   \\
140      & 436 & $\frac{7594}{109}$
\end{tabular}
\caption{The range of interaction $R_m$, the corresponding number of neighbors
 $z$ and effective range of interaction $R$.}
\label{tab:range}
\end{table}

\begin{table}
\begin{tabular}{r|d|d|d|d}
$R_m^2$ & $Q$         & $K_{\rm c}$     & $K_{\rm c}$     & $K_{\rm c}$
                                                      \cite{monbin} \\ \hline
2       & 0.8556 (5)  & 0.1901908 (19)  & 0.1901931 (11)  & 0.190    \\
4       & 0.8557 (9)  & 0.1140216 (18)  & 0.1140225 (7)   & 0.11402  \\
6       & 0.8553 (7)  & 0.0631917 (8)   & 0.0631926 (4)   & \\
8       & 0.8553 (13) & 0.0510460 (10)  & 0.0510467 (4)   & 0.05106  \\
10      & 0.8563 (9)  & 0.0324136 (5)   & 0.03241352 (18) & 0.032463 \\
18      & 0.8555 (14) & 0.0185335 (3)   & 0.01853367 (9)  & \\
32      & 0.853 (3)   & 0.01075152 (25) & 0.01075182 (7)  & \\
50      & 0.856 (6)   & 0.00657274 (26) & 0.00657276 (5)  & \\
72      & 0.854 (4)   & 0.00464056 (16) & 0.00464064 (4)  & \\
100     & 0.850 (8)   & 0.00325903 (15) & 0.00325905 (5)  & \\
140     & 0.862 (17)  & 0.00234637 (19) & 0.00234631 (2)  &
\end{tabular}
\caption{The amplitude ratio $Q$ and critical coupling $K_{\rm c}$ for the
various ranges of interaction studied in this paper. The numbers in
parentheses denote the errors in the last decimal places. The third column
shows the estimates for $K_{\rm c}$ obtained with $Q$ fixed at its Ising value.
For comparison, we also list the estimates for $K_{\rm c}$ given in
Ref.~\protect\cite{monbin}.}
\label{tab:qfit}
\end{table}

\begin{table}
\begin{tabular}{r|d|d|d}
$R_m^2$ & $y_{\rm h}$ & $y_{\rm h}$ & $d_0(R)$    \\ \hline
2       & 1.8745 (7)  & 1.8749 (3)  & 0.9533 (4)  \\
4       & 1.8763 (15) & 1.8756 (4)  & 0.8706 (5)  \\
6       & 1.873 (3)   & 1.8767 (13) & 0.7937 (10) \\
8       & 1.873 (3)   & 1.8754 (8)  & 0.7523 (7)  \\
10      & 1.874 (2)   & 1.8748 (7)  & 0.6783 (6)  \\
18      & 1.871 (3)   & 1.8740 (12) & 0.5816 (6)  \\
32      & 1.875 (6)   & 1.8744 (9)  & 0.4929 (11) \\
50      & 1.873 (7)   & 1.876 (2)   & 0.4181 (18) \\
72      & 1.865 (5)   & 1.8752 (16) & 0.3742 (8)  \\
100     & 1.867 (9)   & 1.877 (2)   & 0.3296 (8)  \\
140     & 1.895 (13)  & 1.879 (3)   & 0.2938 (13)
\end{tabular}
\caption{The exponent $y_{\rm h}$ and the critical amplitude of the
magnetization $d_0(R)$ for the various ranges of interaction. The third column
shows the estimates for $y_{\rm h}$ obtained with $K_{\rm c}$ fixed at the most
accurate values shown in Table~\protect\ref{tab:qfit}.}
\label{tab:mag}
\end{table}

\begin{table}
\begin{tabular}{r|d|d|d}
$R_m^2$ & $y_{\rm h}$ & $y_{\rm h}$ & $p_0(R)$    \\ \hline
2       & 1.8754 (9)  & 1.8748 (2)  & 0.9743 (9)  \\
4       & 1.8753 (12) & 1.8752 (3)  & 0.8136 (7)  \\
6       & 1.8740 (18) & 1.8761 (10) & 0.6762 (14) \\
8       & 1.873 (2)   & 1.8750 (6)  & 0.6076 (9)  \\
10      & 1.874 (3)   & 1.8741 (6)  & 0.4943 (7)  \\
18      & 1.874 (4)   & 1.8740 (11) & 0.3620 (9) \\
32      & 1.868 (4)   & 1.873 (2)   & 0.2622 (9)  \\
50      & 1.862 (6)   & 1.874 (3)   & 0.1914 (7)  \\
72      & 1.863 (17)  & 1.870 (4)   & 0.1534 (8)  \\
100     & 1.870 (6)   & 1.874 (4)   & 0.1180 (8)  \\
140     & 1.86 (3)    & 1.870 (5)   & 0.0954 (9)
\end{tabular}
\caption{The exponent $y_{\rm h}$ and the critical amplitude of the magnetic
susceptibility $p_0(R)$ for the various ranges of interaction. The third column
shows the estimates for $y_{\rm h}$ obtained with $K_{\rm c}$ fixed at the most
accurate values shown in Table~\protect\ref{tab:qfit}.}
\label{tab:chi}
\end{table}

\begin{table}
\begin{tabular}{r|d|d|d}
$R_m^2$ & $y_{\rm h}$ & $y_{\rm h}$ & $v_0(R)$    \\ \hline
2       & 1.8759 (8)  & 1.8754 (3)  & 0.9405 (9)  \\
4       & 1.8744 (12) & 1.8750 (3)  & 0.7860 (6)  \\
6       & 1.8748 (19) & 1.8765 (11) & 0.6528 (15) \\
8       & 1.8746 (17) & 1.8754 (6)  & 0.5862 (9)  \\
10      & 1.875 (3)   & 1.8741 (7)  & 0.4770 (9)  \\
18      & 1.874 (4)   & 1.8745 (10) & 0.3489 (9)  \\
32      & 1.873 (4)   & 1.8747 (18) & 0.2541 (6)  \\
50      & 1.864 (7)   & 1.876 (3)   & 0.1827 (9)  \\
72      & 1.860 (8)   & 1.871 (4)   & 0.1473 (9)  \\
100     & 1.872 (9)   & 1.876 (4)   & 0.1128 (9)  \\
140     & 1.86 (3)    & 1.871 (3)   & 0.0915 (9)
\end{tabular}
\caption{The exponent $y_{\rm h}$ and the critical amplitude of the spin--spin
correlation function $v_0(R)$ for the various ranges of interaction.  The third
column shows the estimates for $y_{\rm h}$ obtained with $K_{\rm c}$ fixed at
the most accurate values shown in Table~\protect\ref{tab:qfit}.}
\label{tab:corr}
\end{table}

\end{document}